\documentclass[12pt]{iopart}
\usepackage{natbib}  
\expandafter\let\csname equation*\endcsname\relax
\expandafter\let\csname endequation*\endcsname\relax
\usepackage{amsmath}
\newcommand{\newblock}{\hskip .11em\@plus.33em\@minus.07em}
\begin{document}
\title[The Big Bang could be anisotropic. The case of Bianchi I model]
{The Big Bang could be anisotropic. The case of Bianchi I model}
          
\author{S L Parnovsky$^{1,2}$}

\address{$^1$Taras Shevchenko National University of Kyiv, Astronomical observatory, Observatorna str. 3, Kyiv 04053, Ukraine}
\address{$^2$CERN 1211 Geneva 23, Switzerland}
\ead{parnovsky@knu.ua}
\vspace{10pt}
\begin{abstract}
We consider an evolution of anisotropic cosmological model on the example of the Bianchi type I homogeneous universe. It is filled by the mixture of matter and dark energy with an arbitrary barotropic equation of state (EoS). The general solution for this case is found and analyzed. A complete list of possible future singularities for this model is given. Some new solution were obtained for a particular EoSs, e.g. for the Bianchi type I $\Lambda$CDM homogeneous model.
It is shown that all special cases corresponding to different EoSs have common properties, provided that now or at another moment of time the Universe is expanding, and the density of the mixture is positive. Then the evolution always begins with an anisotropic``Big Bang'' which happened a finite time ago. After that the universe is constantly expanding and in all cases, with rare exceptions, becomes more isotropic. 
A particularly strong isotropization is associated with the epoch of inflation. After its completion, the expansion of the universe becomes almost isotropic, and this case cannot be distinguished from isotropic by astronomical observations. This fact allows us to consider an anisotropic cosmological model as a possible candidate for the description of the observed Universe despite the isotropic pattern of expansion.
\end{abstract}

%
\vspace{2pc}

\submitto{\CQG}%

\section{Introduction}
For almost 90 years astronomers have known that the Universe is expanding in accordance with Hubble's law 
\begin{equation}
\label{Hub}
v=H r
\end{equation}
for far objects e.g. galaxies. Here $r$ is the distance to the object and $v$ is the velocity of its motion away. 
The Hubble parameter $H$ depends on time, but not on direction. This shows that the modern Universe is isotropic 
on a large scale. Therefore, cosmologists prefer to deal with isotropic homogeneous cosmological models, which reduce to
three Friedmann-Lema\^{i}tre-Robertson-Walker (FLRW) ones and their descendants including the modern $\Lambda$CDM model.

It is known that the homogeneity of the Universe is broken because of the density fluctuations growth. This also leads 
to the appearance of large-scale galaxy collective motion or cosmic flows which yield some deviation from the law (\ref{Hub}). 
However, the assumption that the Universe is and always has been isotropic on a large scale is the standard in astronomy.

We show that anisotropic cosmological models can also agree well with astronomical observations. In this case, the Big Bang was highly anisotropic, but during the inflationary epoch the Universe managed to be isotropized enough that we can not detect a negligible residual anisotropy of cosmological expansion. The currently properties and the future of this model practically coincide with the same for the flat $\Lambda$CDM model. However, the differences in the Big Bang type and the very early stages of the Universe can be significant for some aspects of a present picture of the world.

We demonstrate this possibility using the anisotropic homogeneous cosmological model Bianchi type I \citep[see the 
Bianchi classification e.g. in $\S$112 of][]{LL}. Using the homogeneous model we simplifies the analysis, reducing the system of partial differential equations to to the system of ordinary differential equations. 

The model of Bianchi type I  is the simplest anisotropic one. In addition, it is spatially flat, which corresponds to the well-known fact that our Universe is spatially flat or almost spatially flat.

As a by-product, we solve the problem, which is enough interesting for the relativistic cosmology. We find and analyse the generic solution for the cosmological model Bianchi type I with matter with arbitrary barotropic equation of state (EoS).

All exact solutions of Einstein equations are valuable in general relativity. They enable a study of the properties 
of the corresponding space-times. Some solutions underlie the modern astrophysical and cosmological conceptions. 
There are hundreds of exact solutions in general relativity describing e.g. in the book \citep{Mac}.

So many solutions are because one can consider the different structure and type of a space-time, as well as a matter 
with a specific equation of state. Scientists consider solutions with vacuum, cosmological constant, electromagnetic, scalar, vector, spinor or tensor fields, light radiation, dust matter, matter with pressure, dark energy of different 
kinds and combinations of mention above factors. Therefore, the exact solutions with enough general energy-momentum 
tensor, which can replace a number of such solutions, are of particular interest. 

In this paper we consider the Bianchi type I model for the case the Universe is filled by all kinds of matter and dark energy with a total mass density $\rho$ and an effective pressure $p$. 
The barotropic equation of state $p(\rho)$ is general enough to include all the particular cases most interesting for cosmology.
Nevertheless we can describe the Bianchi type I model with such matter by a single metric. In the particular cases of
certain $p(\rho)$ it reduces to some well-known metrics. It is demonstrated in \Sref{part}. Some new cases are studied 
in \Sref{new}.

The relations between the mass density $\rho$, cosmological time $t$ and the parameters of the Universe (size and 
analogue of the Hubble parameter) is obtained as integrals. This will enable the cosmologists and astronomers to 
deal with this solution without solving the Einstein equations directly as well as they use the Friedmann equations 
considering the FLRW homogeneous isotropic cosmological model.

\section{Solution of the Einstein equations for the Bianchi type I cosmological model}

The metric of the homogeneous cosmological model of the Bianchi type I in the synchronous frame has the form 
\begin{equation}
\label{e1}
\rmd s^{2} = \rmd t^{2} - \gamma_{ik}(t)\rmd x^i \rmd x^k\quad(i=1,2,3).
\end{equation}
We use the same definitions and notations as in a well-known book \citep{LL}. In particular, the
signature ($+---$) is used. We use the system of units in which $G = 1$ and $c = 1$, so
the energy density $\varepsilon$ coincides with $\rho$. We use Roman letters as the indices for spacial coordinates 
in contrast to \citep{LL} who used Greek ones. The Greek indices label the four-dimensional coordinates.
The coordinate $t=x^0$ is time-like one, so we exclude the solutions
describing strong gravitational waves \citep{B}, naked singularities \citep{HP, P80} and other space-times which are 
not interesting for cosmology \citep{P79}. We will use the notation $x^{1} = x,\; x^{2} = y,\; x^{3} = z$.

This space-time is filled with a homogeneous matter with a mass density $\rho(t)$ and a
pressure $p(\rho)$. The synchronous frame of reference is also the co-moving one and the energy-momentum
tensor is diagonal in this frame $T_{\alpha\beta}=diag(\rho,-p,-p,-p)=U_{(\alpha)}\delta_{\alpha\beta}$ (no summation over the
$\alpha$ index in all equations with $U_{(\alpha)}$). 
The three-dimensional spatial Ricci tensor $P_{ab}$ vanishes for Bianchi type I model, so 
the spatial components of Einstein equations give us the relations \citep{LL}
\begin{equation}
\label{e2}
R_{\alpha}^{\beta}=-\frac{1}{2\sqrt{\gamma}}\frac{\rmd}{\rmd t} (\sqrt{\gamma}\kappa_{\alpha}^{\beta})=
8\pi\left(T_{\alpha}^{\beta}-\frac{T}{2}\delta_{\alpha}^{\beta}\right),
\quad\gamma=det|\gamma_{\alpha\beta}|,\quad\kappa_{\alpha}^{\beta}=\frac{\rmd \gamma_{\alpha\gamma}}{\rmd t}\gamma^{\beta\gamma}.
\end{equation}
The density $\rho$ and therefore the pressure $p$ depend only on cosmological time $t$. From (\ref{e2}) we get
\begin{equation}
\label{e3}
\kappa_{\alpha\beta}=\frac{1}{2\sqrt{\gamma}}\lambda_{\alpha\beta}-16\pi\delta_{\alpha\beta}\int (2U_{(\alpha)}-T)\rmd t,
\end{equation}
where $\lambda_{\alpha\beta}$ is a constant tensor. We can diagonalize it by transformation of the coordinate system \citep[see details in $\S$103 of textbook by][]{LL}. The tensors $\kappa_{ik}$ and $\gamma_{ik}$ also get a diagonal form.
So, we consider the metric of the homogeneous cosmological model of the Bianchi I type in the diagonal form
\begin{equation}
\label{eq1}
\rmd s^{2} = \rmd t^{2} - \rme^{2h_1(t)}\rmd x^{2} - \rme^{2h_2(t)}\rmd y^{2} - \rme^{2h_3
(t)}\rmd z^{2}
\end{equation}
without loss of generality.

Let`s consider the Einstein equations.   
All their non-diagonal components are satisfied and the
diagonal components give the following conditions 
\begin{equation}
\label{eq2}
 - R_{i}^{i} = \ddot {h_i}  + \dot {h_i} (\dot {h_1}  + \dot {h_2} 
+ \dot {h_3} ) = 4\pi (\rho-p),
\end{equation}
\begin{equation}
\label{eq3}
 - R_{0}^{0} = \ddot {h_1}  + \ddot {h_2}  + \ddot {h_3}  + \dot
{h_1} ^{2} + \dot {h_2} ^{2} + \dot {h_3} ^{2} =- 4\pi (\rho+3p),
\end{equation}
where a dot denotes time derivative. Note that there is no summation over $i$ in the 
left side of the equations (\ref{eq2}).

Subtracting equations (\ref{eq2}) from each other we get
after integration
\begin{equation}
\label{eq4}
\dot {h_1}  + \dot {h_1}  + \dot {h_1} = \mathrm{const} - \ln (\dot {h_1} 
- \dot {h_2} ) = \mathrm{const} - \ln (\dot {h_1}  - \dot {h_3} )
 = \mathrm{const} -\ln (\dot {h_2}  - \dot {h_3} ).
\end{equation}
This leads us to
\begin{equation}
\label{eq5}
\dot {h_i}  = \dot{f} + q_{i} \rme^{ - 3f}\quad(i=1,2,3),
\end{equation}
where $q_i$ are constant quantities having the dimension s$^{-1}$, $3f = h_1 + h_2 + h_3$ and
\begin{equation}
\label{eq6}
q_{1} + q_{2} + q_{3} = 0.
\end{equation}

The quantity $\dot{f}$ has a simple physical interpretation. If we denote by $V$ the volume of the 
part of the Universe, which borders have fixed spatial coordinates $x,\, y,\, z$, we get 
$3\dot{f}=d(\ln(V))/dt$. In the isotropic case $\dot{f}$ corresponds to the Hubble parameter $H$ and the
quantity $f$ to the natural logarithm of the scale factor $\ln(a)$.

The values $\dot {h_i}$ characterize the time derivatives of the logarithms of the scale factors along the $x,\,y,\,z$ 
axes. They play a role of ``Hubble parameters'' along three axis of this anisotropic model. They are the sum of the general term $\dot{f}$ corresponding to the isotropic expansion and the anisotropic term equal to the product of $\rme^{ - 3f}$ by the coefficient $q_{i}$. So, we can introduce the value
\begin{equation}\label{an}
W= \rme^{ - 3f}/3\dot{f}
\end{equation}
to describe quantitatively an anisotropy of the model. The factor $1/3$ is introduced in order that $W=1$ at the anisotropic ``Big Bang''. Decreasing of $W$ means isotropization. We consider this process in \Sref{is}.

Two combinations of (\ref{eq2}) and (\ref{eq3}) are especially useful, namely
\begin{equation}\label{eq8}
\frac{\rmd\rho}{\rmd t}=-3(\rho+p)\dot{f},
\end{equation}
which describes the energy conservation law and
\begin{equation}\label{eq9}
8 \pi \rho=3\dot{f}^2-\frac{Q}{2}\rme^{-6f}
\end{equation}
with the constant
\begin{equation}\label{eq7}
Q=q_1^2+q_2^2+q_3^2 \geq 0.
\end{equation}

In the isotropic case $h_1=h_2=h_3$ and $q_1=q_2=q_3=Q=0$ and we deal with the flat FLRW metric, possibly with 
the addition of a dark energy of some kind. 
In this case (\ref{eq9}) means $\dot{f}=H=(8\pi \rho /3)^{1/2}$.
This trivial case is well studied. One can find the complete analysis in \citep{P15}.
So, hereafter we will only consider the anisotropic case with $Q>0$.

If $\rho+p=0$
we must put $\rho=const$. This case of the pure cosmological constant will be considered in \Sref{part}.
If $\rho+p\neq 0$ we obtain from (\ref{eq8})
\begin{equation}\label{eq10}
f=-\frac{1}{3}\int\frac{\rmd\rho}{\rho+p(\rho)}.
\end{equation}
So, we can find the dependence $f(\rho)$ and therefore the inverse relationship $\rho(f)$. 
We can substitute $\rho(f)$ into (\ref{eq9}) and calculate
\begin{equation}\label{eq11}
t=\int\left(\frac{8\pi}{3}\rho+\frac{Q}{6}\rme^{-6f}\right)^{-1/2}\rmd f,
\end{equation}
obtaining the relationship $t(f)$ and the inverse relationship $f(t)$. Thus, we get the solution of the Einstein equations for our case.

Moreover, after switching to the new time variable $f$ we can present the metric which realize (\ref{eq1}) in this case
\begin{equation}
\label{eq12}
\rmd s^2=\frac{6\rmd f^2}{16\pi \rho +Qe^{-6f}}-
\rme^{2f}\left( \rme^{2q_1 S(f)}\rmd x^2+\rme^{2q_2 S(f)}\rmd y^2+\rme^{2q_3 S(f)}\rmd z^2 \right),
\end{equation}
\begin{equation}
\label{eq13}
S=\int \left( \frac{8\pi}{3} \rho \rme^{6f}+\frac{Q}{6}\right) ^{-1/2}\rmd f.
\end{equation}
Together with (\ref{eq10}) this gives the generic solution of the Einstein equations in the case of the Bianchi type I space-time filled with the matter with the arbitrary barotropic equation of state
$p(\rho)$.

Note that equations \eref{eq10} and \eref{eq11} give any researcher an opportunity to study the dependence
of the parameters of the model without using the mathematical apparatus of the GTR. One can say that they play 
the role of Friedmann equations for the isotropic FLRW models.

\section{Some known particular solutions} \label{part}

If matter is absent and $\rho=p=0$ the solution (\ref{eq12},\ref{eq13}) reduces to the well-known
Kasner metric \citep{Kas}
\begin{equation}
\label{eq14}
\rmd s^{2} = \rmd t^{2} - t^{2p_{1}} \rmd x^{2} - t^{2p_{2}} \rmd y^{2} - t^{2p_{3}}\rmd z^{2},
\end{equation}
where dimensionless indices $p_{i}$ are connected with the $q_i$ values by
\begin{equation}
\label{eq15}
p_{i}=\frac{1}{3}+\left( \frac{2}{3Q}\right) ^{1/2}q_i.
\end{equation}
The Kasner indices satisfy the conditions
\begin{equation}
\label{eq16}
p_{1} + p_{2} + p_{3} = 1,\quad {p_{1}}^{2}
+{p_{2}}^{2} +{p_{3}}^{2} = 1.
\end{equation}

When we reduce the metric to the form (\ref{eq14}), we choose a specific scale along the $x,y,z$ axes. As a result we have $W=1$ and $Q=2/3\, s^{-2}$ for the metric (\ref{eq14}).

The solution becomes the Heckmann-Schuking one \citep{HS} if the dust matter is present with $p=0,\rho=\rho_0\exp(-3f)\neq0$ (it is described by (\ref{eq17}) with $w=0$). The metric is
\begin{equation}
\label{e4}
\rmd s^{2} = \rmd t^{2} - a_0^{\phantom{0}2}t^{2p_{1}}\tau^{4/3-2p_{1}} \rmd x^{2} - b_0^{\phantom{0}2}t^{2p_{2}}
\tau^{4/3-2p_{2}} \rmd y^{2} - c_0^{\phantom{0}2}t^{2p_{3}}\tau^{4/3-2p_{3}}\rmd z^{2},\quad\tau=t+t_0.
\end{equation}
The constant $t_0$ can be expressed by formula
\begin{equation}
\label{e4a}
t_0=\frac{2Q^{1/2}}{3\rho_0}.
\end{equation}
Note that in system of units we use, $Q$ and $\rho_0$ are measured in s$^{-2}$. An isotropization in this model is well-known. 

At $p=-\rho$ we deal with the cosmological constant $\Lambda=8\pi \rho=\mathrm{const}>0$.
In this case the metric after some re-scaling has the form
\begin{align}
\label{eq18}
& \rmd s^2 = \rmd t^2 - \cosh ^{4/3}(\lambda t)\left( u^{2p_1} \rmd x^2+u^{2p_2} \rmd y^2  
 +u^{2p_3}\rmd z^2 \right), \nonumber\\
&u= \lambda^{-1}\tanh(\lambda t),\quad \lambda = \frac{(3\Lambda)^{1/2}}{2}.
\end{align}
This space-time has only one singularity at $t = 0$. It has the Kasner form
(\ref{eq13}). At large $t$ (\ref{eq18}) tends to the well-known de Sitter metric. It also shows an apparent isotropization.

To make the picture complete we can consider the case $\Lambda<0$, which is popular in general relativity, 
although speculative. In this theoretical case
\begin{align}
\label{eq19}
& \rmd s^{2} = \rmd t^{2} - \cos ^{4 / 3}(\lambda t)\left( u^{2p_1} \rmd x^2+u^{2p_2} \rmd y^2  
 +u^{2p_3}\rmd z^2 \right), \nonumber\\
&u= \lambda^{-1}\tan(\lambda t),\quad \lambda = \frac{(-3\Lambda)^{1/2}}{2}.
\end{align}
This space-time has a Kasner singularity at $t = 0$, corresponding to the
birth of the Universe, and another singularity at $t_0 = \pi (-3\Lambda)^{-1/2}$, 
corresponding to the ``Big Crunch''. If $t \to t_0$, the
metric (\ref{eq19}) reaches its asymptotic form, which can be obtained from (\ref{eq14}) by
substituting $t$ by $t_0 - t$, $p_{i} $ by $\tilde{p}_{i} = 2 / 3 - p_{i} $
and rescaling the spatial coordinates. One can see from (\ref{eq15}) that $\tilde{p}_{i}$ 
is also a set of Kasner indices. So, both initial and final
singularities are Kasner ones. There is no isotropization in this case.


\section{Some new metrics}\label{new}
\subsection{$\Lambda$CDM solution for Bianchi type I cosmological model} \label{lcdm}

Let us show how it is easy to obtain an interesting special case from the general equation. We start with an example of the $\Lambda$CDM model, which is usually associated with the FLRW metric. However, this designation means that the universe is filled with cold dark matter (CDM), the pressure of which can be neglected and dark energy acting as a pure cosmological constant. Let us consider a homogeneous Bianchi type I cosmological model with the energy-momentum tensor being the sum of the tensors for these two components. For the cosmological constant we have $\rho_{\Lambda}=-p_{\Lambda}=\Lambda/8\pi=\mathrm{const}>0$ and for CDM we have $\rho_{CDM}\propto V^{-1}\propto \rme^{-3f}=A\rme^{-3f}$ with $A=\mathrm{const}$, $p_{CDM}=0$. Here $V$ is the volume occupied by the constant mass of CDM, which increases as the universe expands. From (\ref{eq11}) we get
\begin{eqnarray}\label{lcdm1}
t&=&\int\left(\frac{\Lambda}{3}+\frac{8\pi A}{3}\rme^{-3f}+\frac{Q}{6}\rme^{-6f}\right)^{-1/2}\rmd f \nonumber\\
&=&(3\Lambda)^{-1/2}\ln \left[ 2\sqrt{\Lambda(\Lambda\rme^{6f}+8\pi A \rme^{6f}+Q/2)}+2\Lambda\rme^{3f}+8\pi A \right] + t_0.
\end{eqnarray}
Here $\Lambda>0$ and $t_0=\mathrm{const}$. If $Q\Lambda > 32\pi^2 A^2$ we get
\begin{equation}\label{lcdm2}
2\Lambda\rme^{3f}=\sqrt{2Q\Lambda - 64\pi^2 A^2}\,\sinh{\left[\sqrt{3\Lambda} (t-t_0)\right] -8\pi A}.
\end{equation}
The solution for the case $\Lambda<0$ has the form
\begin{equation}\label{lcdm3}
2\Lambda\rme^{3f}=\sqrt{64\pi^2 A^2- 2Q\Lambda }\,\sin{\left[\sqrt{3|\Lambda|} (t_0-t)\right] -8\pi A}.
\end{equation}
We can easily include the radiation into this model, consider the dark energy with different equation of state instead of cosmological constant, etc.
\subsection{Other examples of new solutions} \label{new2}
Consider the model with the matter
satisfying the most popular EoS  
\begin{equation}
\label{eos1}
p=w\rho
\end{equation} 
with $w=$const. In this case 
we yield from (\ref{eq10})
\begin{equation}
\label{eq17}
\rho=\rho_0 \rme^{-3(1+w)f}, \quad \rho_0=\mathrm{const}.
\end{equation}
After substitution (\ref{eq17}) into (\ref{eq11}) we see that at $w<-1$ i.e. for phantom energy the integral 
is finite even if the upper limit of integration becomes infinite. So, $f \xrightarrow [t \to t_0]{}\infty$. 
This means that volume $V$ becomes infinite in finite time. 
This is the ``Big Rip'' case and it occurs at some time point $t_0$ \citep{CKW}.
One can consider the asymptotic behaviour of the Universe immediately before the ``Big Rip'' i.e. at small
$\Delta t=t_0-t$ and get $f \xrightarrow [t \to t_0]{}2\ln(\Delta t)/[3(1+w)]+\mathrm{const}$.

We can apply other barotropic equations of state to (\ref{eq10}). For example $p=w(\rho-\rho_0)$ used by \citet{H} and
\citet{BBQ}, $p=p_0+\alpha\rho+\beta\rho^2$ used by \citet{A} or $p=\gamma\rho/(1-\beta\rho)-\alpha\rho^2$ \citep{Cap}. 
These equations allow $p(\rho)$ to cross the line $p=-\rho$ between the domains of the ordinary and the phantom energies. We will consider this possibility in \Sref{cross}. Note that the integral (\ref{eq10}) can be taken analytically for many equations of state, which are considered as cosmological models or as toy models, e.g.
for generalized Chaplygin gas.

\subsection{Not so directly barotropic equation of state} \label{nondirect}
In many cases cosmologists use equations of state depending from the scale factor of the Universe $a$ or the redshift $z=a_0/a(t)-1$, where $a_0$ is the scale factor at the present time. For the Bianchi type I model we have to use $\exp(f)$ instead of $a$ in this equations. Many EoSs have the form
\begin{equation}
\label{e5}
p+\rho=F(\rho)E(f),
\end{equation}
usually
\begin{equation}
\label{e6}
p=w(a)\rho,\qquad a=a_0\exp(f-f_0),
\end{equation}
where $f_0$ is the value of $f$ at present time.
In this cases (\ref{eq8}) provides the relation between $f$ and $\rho$ in the form
\begin{equation}
\label{e7}
\int F(\rho)^{-1}\rmd \rho=-3\int E(f)\rmd f.
\end{equation}
After substitution $\rho(f)$ into (\ref{eq12},\ref{eq13}) we get the metrics for these models.

There are a lot of equations of state in the form (\ref{e6}) from the popular among cosmologists phenomenological one $w=w_0+w_1z$ to $w=(1+z^{1/\tau})^{-1}$ for ``aether'' \citep{BCL}. In this case one get 
\begin{equation}
\label{e8}
\rho=\exp\left(-3\int(w(a)+1)a^{-1}\rmd a\right).
\end{equation}
For example, equation $w=w_0+w_1z$ provides
\begin{equation}
\label{e9}
\rho=\exp\left[-3\left( (1+w_0-w_1)f-w_1\exp(f_0-f)\right)\right].
\end{equation}

\section{Analysis of common properties of the Bianchi type I cosmological models} \label{an0}

Let us analyze the main properties of the space-time with the metric (\ref{eq12},\ref{eq13}). We will use
two additional facts, namely, that at the present time $\rho\geq0$ and that the Universe is expanding, which
corresponds to $\dot{f}>0$.

\subsection{Only expansion} 

From the \eref{eq9} we see that the stop of expansion $\dot{f}=0$ can occur only if $\rho=0$ and $f\to \infty$. So the
Universe will expand permanently. This can be an eternal expansion or an expansion till the ``Big Rip'' or other
future singularity. Expansion can stop only if effective density $\rho<0$, which corresponds e.g. to the case of 
negative cosmological constant $\Lambda$. From \eref{eq19} one can see that in this case the Universe begin to collapse till the ``Big Crunch'' which has the Kasner type.

\subsection{``Big Bang''} 

Looking in the past we see that expansion was not eternal and it have to start from the ``Big Bang'', 
corresponding to $f\to -\infty$. From \eref{eq11} we can obtain a restriction on the age of the Universe
$T$ within the considered model
\begin{equation}\label{eq20}
T=\int_{-\infty}^{f_0}\left(\frac{8\pi}{3}\rho+\frac{Q}{6}\rme^{-6f}\right)^{-1/2}\rmd f \leq \sqrt{\frac{6}{Q}}
\int_{-\infty}^{f_0}\rme^{3f}\rmd f= \sqrt{\frac{2}{3Q}} \exp(3f_0).
\end{equation}
As a result, the model requires the ``Big Bang'' in the past.
 
Let us investigate the type and properties of this initial cosmological singularity. Suppose that the Universe is
filled with the mixture of different kinds of matter, labeled by subscript $a$. All types of matter have the equations 
of state $p_a=w_a\rho_a$ with different $w_a=\mathrm{const}$ and their dependencies on $f$ have the form \eref{eq17}. If all
$w_a<1$ then at $f\to -\infty$ the term with $\rho$ in \eref{eq11} and the metric (\ref{eq12}) becomes a negligible correction to the main term $\propto \exp(-6f)$. The matter asymptotically fits the vacuum metric (\ref{eq16}) near the ``Big Bang'' without affecting the Kasner type singularity. Note that the dust matter, cold or warm dark matter, 
the radiation, the cosmological constant refer to this case. The ordinary matter and radiation must satisfy
the condition $T_i^i\geq 0$ \citep[see $\S$34 of][]{LL} which means $w_a\leq 1/3$ if $\rho>0$.

We can consider theoretically some unknown kind of matter or energy violating this condition. If $w_a<1$ it cannot change the solution near the ``Big Bang''. But if there is the exotic matter with $w_a>1$ among other kinds of matter, it would change the asymptotic behaviour of metric at $f\to -\infty$ to $f=2\ln(t)/[3(1+w_a)]+\mathrm{const}$. Here the moment of time $t=0$ corresponds to singularity.
From \eref{eq13} we have $S\propto t^u$ with $u=(1-w)/(1+w)<0$. Solution in this case is anisotropic.

Note, that if the Universe is filled with such exotic matter with $w>1$ only then the model will expand forever
and the matter asymptotically fits the vacuum metric (\ref{eq16}) at $f\to \infty$. The presence of such a matter 
will not affect the model on the late phase of its evolution because of $\rho \ll \mathcal{O}(\exp(-6f))\to 0$.

\subsection{Possible future singularities}\label{future}

We see that the Universe will expand forever or this process will last a finite time and end in a singularity, which is called a future singularity. For the matter with 
EoS \eqref{eos1} with $w<-1$ a ``Big Rip'' is inevitable. Other types of future singularities are also possible. 
Their classification was carried out in the paper \citep{NOT} for the isotropic FLRW model. Four possible types were found 
for the singularities at $t=t_0$ with finite $t_0$. They include:
\begin{description}
\item[$\bullet$ Type I] $a,\rho,|p|\to \infty$ (``Big Rip'')
\item[$\bullet$ Type II] $a\to a_0;\rho \to \rho_0;|p|\to \infty$ (``sudden singularities'')
\item[$\bullet$ Type III] $a\to a_0;\rho,|p|\to \infty$ (it was named ``Big Freeze'' in \citep{BGH})
\item[$\bullet$ Type IV] $a\to a_0;\rho,|p|\to 0$ and higher derivatives of the Hubble parameter $H$ diverge.
\end{description}
In the case of anisotropic Bianchi I cosmological model we have to use $f$ and $\dot{f}$ instead of $a$ and $H$.

There are some singularities with $t_0=\infty$, too. 
The ``Little Rip'' singularity \citep{FLS}, similar to the ``Big Rip'', but with an eternal expansion is among them. 

According to (\ref{eq10}) all nontrivial solutions arise if $\rho+p \to 0$ or $\rho+p \to \infty$. The list of possible future singularities for isotropic FLRW model was prepared in \citep{P15}. We can use some EoS of matter near the singularities of different types from this paper. 

Let us briefly describe the results obtained. If singularity occurs at $\rho \to \infty$ we can consider either the EoS \eqref{eos1} with $w<-1$ or the asymptotic form of EoS
\begin{equation}\label{eos2}
\rho+p\xrightarrow [\rho \to \infty]{}-A\rho^\alpha
\end{equation}
with $A,\alpha=\mathrm{const}$. The latter corresponds to the case when $w$ is not constant, but asymptotically tends to $-1$ at $\rho/p\xrightarrow [\rho \to \infty]{}-1$. In this case $\alpha<1$. The consideration of more complex asymptotic forms of EoS gives nothing new in comparison with this EoS. If $\alpha<1$ then $f\propto \rho^{1-\alpha}\xrightarrow [\rho \to \infty]{}\infty$, $t\to t_0+\mathrm{const}\cdot f^{(2\alpha-1)/(2\alpha-2)}$. If $1/2<\alpha<1$ then $t
\xrightarrow [\rho \to \infty]{}t_0$ and we deal with the ``Big Rip''. At $\alpha<1/2$ the integral (\ref{eq11}) diverges and we deal with the ``Little Rip'' case. If $\alpha>1$ we deal with the ``Big Freeze'' with $f\xrightarrow [\rho \to \infty]{}f_0+\mathrm{const}\cdot \rho^{1-\alpha}$, 
$t\xrightarrow [\rho \to \infty]{}t_0$. 

Thus, the future singularities with $\rho \to \infty$ for the anisotropic cosmological Bianchi type I model are the same as for isotropic FLRW model. It is easy to prove that the same situation occurs for the sudden singularities with $\rho \to \rho_0\neq 0$ and the equation of state 
\begin{equation}\label{eos3}
\rho+p(\rho)\xrightarrow [\rho \to \rho_0]{}C(\rho-\rho_0)^\mu
\end{equation}
with $\mu,C=\mathrm{const}$. 

To finish the study, consider the last possibility of zero terminal density with the  power-law asymptote of the EoS
\begin{equation}\label{eos4}
\rho+p\xrightarrow [\rho \to 0]{}-D\rho^\nu
\end{equation}
Substituting it into (\ref{eq10}) we obtain
$f\xrightarrow [\rho \to 0]{}f_0+\mathrm{const}\cdot \rho^{1-\nu}$. If $\nu>1$ then $f\xrightarrow [\rho \to 0]{}\infty$.
This case corresponds to the ``Little Freeze'' case introduced in \citep{P15} and has the same asymptotes. If $\nu<1$ then 
$f\xrightarrow [\rho \to 0]{}f_0=\mathrm{const}$. This is the case in which the difference between the properties of the isotropic and anisotropic models becomes essential.
The second term in brackets in the integral (\ref{eq11}) is proportional to $Q$ and arise due to the anisotropy
of the model. In this case it is the main term and the term $\propto \rho\, \exp(6f)$, caused by the matter, is negligible. 

As a result, $t\xrightarrow [\rho \to 0]{}t_0$. No ``Little Freeze'' can exist if $1/2<\nu<1$, it corresponds to the type IV singularity. The ``Big Squeeze'' also introduced in \citep{P15} occurs at $\nu<0$, at $0<\nu<1/2$ we still deal with the type IV singularity. However, the asymptotic behavior of the anisotropic model differ from the isotropic one.
Instead of asymptotes listed in \citep{P15} we get $\rho \propto \Delta t^{1/(1-\nu)}$ and $f \propto \Delta t$. 
The list of possible future singularities for Bianchi I cosmological model is presented in Table \ref{t1}. The asymptotic behaviour of parameters near future singularities are listed in Table \ref{t2}. There are a few minor differences from the isotropic case. 

Let me clarify that the existence of such singularities is not obligatory in Bianchi type I models. They are considered in this section because they are mentioned in the next subsection, which discusses the extremely important process of isotropization in the evolution of the universe, described by a metric of Bianchi type I.

\begin{table*}[tb]
\caption{List of possible future singularities for Bianchi I cosmological model}
\begin{tabular}{llcccccc} 
\hline
\hline
\textnumero&Type&Nickname&EoS&$\rho_0$&$|p_0|$&$p_0+\rho_0$&$f_0$\\
\hline
\multicolumn{8}{c}{$t \to t_0$, $\Delta t=t_0-t\to 0$} \\ \hline
1&I& ``Big Rip''&(\ref{eos1}), $w<-1$&$\infty$&$\infty$&$-\infty$&$\infty$\\
2&I& ``Big Rip''&(\ref{eos2}), $1/2<\alpha<1$&$\infty$&$\infty$&$-\infty$&$\infty$\\
3&III&``Big Freeze''&(\ref{eos2}), $\alpha>1$&$\infty$&$\infty$&$-\infty$&$f_0$\\
4&II& ``sudden''&(\ref{eos3}), $\mu<0$&$\rho_0$&$\infty$&$-\infty$&$f_0$\\
5&IV&&(\ref{eos4}), $0<\nu<1$&$0$&$0$&$0$&$f_0$\\
6&IV&``Big Squeeze''&(\ref{eos4}), $\nu<0$&$0$&$\infty$&$-\infty$&$f_0$\\
\hline
\multicolumn{8}{c}{$t \to \infty$} \\ \hline
7&& ``Little Rip''&(\ref{eos2}), $0<\alpha<1/2$&$\infty$&$\infty$&$-\infty$&$\infty$\\
8&& ``Little Rip''&(\ref{eos2}), $\alpha<0$&$\infty$&$\infty$&$0$&$\infty$\\
\hline \hline

\end{tabular}\label{t1}
\end{table*}
\begin{table*}[tb]
\caption{Asymptotic behaviour near future singularities}
\begin{tabular}{lccc}
\hline
\hline
\textnumero&$\rho$&$p+\rho$&$f$\\
\hline
1&$\propto \Delta t^{-2}$&$\propto \Delta t^{-2}$&$f\to\mathrm{const}-2/(3|1+w|)\ln(\Delta t)$\\
2&$\propto \Delta t^{2/(1-2\alpha)}$&$\propto \Delta t^{2\alpha/(1-2\alpha)}$
&$f\propto \Delta t^{2(1-\alpha)/(1-2\alpha)}$\\
3&$\propto \Delta t^{2/(1-2\alpha)}$&$\propto \Delta t^{2\alpha/(1-2\alpha)}$
&$f\to f_0-\mathrm{const}\Delta t^{(2\alpha-2)/(2\alpha-1)}$\\
4&$\rho-\rho_0\propto \Delta t^{1/(1-\mu)}$&$\propto \Delta t^{\mu/(1-\mu)}$
&$f\to f_0-\dot{f}_0\Delta t$\\
5&$\propto \Delta t^{1/(1-\nu)}$&$\propto \Delta t^{\nu/(1-\nu)}$&$f\to f_0-\dot{f}_0\Delta t$\\
6&$\propto \Delta t^{1/(1-\nu)}$&$\propto \Delta t^{\nu/(1-\nu)}$&$f\to f_0-\dot{f}_0\Delta t$\\
7&$\propto t^{2/(1-2\alpha)}$&$\propto t^{2\alpha/(1-2\alpha)}$&
$f\propto t^{2(1-\alpha)/(1-2\alpha)}$\\
8&$\propto t^{2/(1-2\alpha)}$&$\propto t^{2\alpha/(1-2\alpha)}$&$f\propto t^{2(1-\alpha)/(1-2\alpha)}$\\
\hline \hline

\end{tabular}\label{t2}
\end{table*}
\subsection{Isotropization}\label{is}

Taking into account \eqref{eq9} we get for the value \eqref{an} characterised the anisotropy
\begin{equation}\label{an1}
W= \left(24\pi\rho\rme^{6f}+1.5Q\right)^{-1/2}.
\end{equation}
If we choose scales along the spatial axes in such a way that near the ``Big Bang'' the metric has the form like the Kasner
one, we have $Q=2/3$ and $W=1$ at $\rho\rme^{6f}\to 0$. 
 
The isotropisation requires $\rho\rme^{6f}\to\infty$. It is possible in the cases of $f\to\infty$ corresponding to the eternal expansion up to an infinite scale factor or the ``Big Rip''. Another possibility is the ``Big Freeze'' with $\rho\to\infty$. 
Naturally, one can come up with a model in which
a matter with a special EoS expands eternally without isotropisation due to a very strong decrease in density upon expansion, 
but this is purely speculative. A presence of a dust-like matter or radiation excludes this possibility.

Let's make some quantitative estimates of the change of the $W$ value during the evolution of the model. We focus on its application to the real Universe. An evolution started from the anisotropic ``Big Bang'' similar to the Kasner metric \eqref{eq14} with $W=1$.
Immediately or almost immediately, inflationary expansion began. It was caused by a short-term factor acting like a large effective cosmological constant. Therefore, we can use the metric \eqref{eq18} with 
\begin{equation}\label{an2}
W= \frac{1}{1+2\sinh^2\lambda t}
\end{equation}
to describe this stage. If the Universe expanded by a factor of $K$, this value decreased by approximately $K^{3}/2$ times. Using the 
value $K=10^{26}$ we estimate the degree of anisotropy $W$ at the level of $10^{-78}$ after inflation. This value is only reduced upon subsequent expansion within the framework of the $\Lambda$CDM model. Obviously, at present we can not observe such a small residual anisotropy of the Hubble expansion. It was almost isotropic even at the end of the inflationary epoch. We see that due to the strongest isotropization at the inflationary stage, the anisotropic universe becomes practically isotropic. 
 
\subsection{Crossing the border between domains of an ordinary matter and a phantom one}\label{cross}

If the Universe is filled with a substance with an exotic EoS, corresponding to either ordinary or phantom matter depending on its density, it could cross the border $p=-\rho$ between these domains when density changes. Is it possible? 

If the intersection occurs at a finite density $\rho_0$, then the EoS near it has an asymptotic form \eqref{eos3} with $\mu>0$. If $0<\mu<1$ this corresponds to some finite value $f=f_0$ and a finite time $t_0$ according to \eqref{eq10} and \eqref{eq11}. So the intersection is possible and it occurs in a finite time interval. 

We can go further and consider an even more exotic case with two points of intersection of the line $p=-\rho$ and the EoS $p(\rho)$ with asymptotic form  near both intersection points if form \eqref{eos3} with $0<\mu<1$. In this case a closed trajectory in $p,\rho$ plane that intersects the line $p=-\rho$ at precisely these two points is possible, despite the Universe expands and $f$ increases constantly. The transition from the ordinary matter domain to phantom one occurs at a lower density. While the parameters of matter correspond to the phantom domain, its density increases. Then it crosses the border at higher density, falls into the domain of an ordinary matter and its density decreases. 

If the system then returns to the starting point on the $p,\rho$ plane, i.e. the point of the first intersection of the boundary of two domains, then in the future it will repeat its path in cycles with a constant period of $f$. This can be seen from \eqref{eq8} rewritten as
\begin{equation}\label{eq8a}
\frac{\rmd\rho}{\rmd f}=-3(\rho+p).
\end{equation} 
As a result, we get a cosmological model looking like a solution with changing positive cosmological constant, which is a periodic function of $f$. However, its cosmological time period increases according to \eqref{eq11}, tending to some constant value.

What is the basis for the assumption that the trajectory of the system is closed? Let us assume that there are only two points of its possible intersection on the line $p=-\rho$, which we will call points 1 and 2. Point 1 has a greater density than 2. Consider the trajectory of the system described by the Einstein equations with matter with barotropic EoS on the $p,\rho$ plane. Let the system pass from the phantom domain to the ordinary one at point 1, make some path to point 2 inside the domain of ordinary matter and return to the phantom domain at point 2. Then, most likely, the further trajectory of the system inside the phantom domain will lead it back to point 1.

Consider two trajectories of the system inside the phantom domain. The first one is the trajectory after crossing the boundary at point 2, while moving along it, the density increases. The second is the trajectory along which the system got to point 1, but after the inversion of the arrow of time direction. When moving along it, the density decreases. Obviously, these trajectories must intersect at some point. This cannot be a regular point of a dynamical system. There are two possibilities left. Either the point of intersection is a singular point of the system lying outside the line $p=-\rho$, which have to be a sudden singularity according to Table \ref{t1} (this case requires a very specific EoS for matter and dark energy), or the two trajectories simply coincide. The last option corresponds to the evolution of the system, automatically returning it from point 2 to point 1. The same thing happens in the first case, but then the trajectory of the system passes through a sudden singularity with $p\to -\infty$.

At $\mu>1$ we get $f_0=t_0=\infty$. This corresponds to the eternal expansion with an asymptotic approximation of EoS 
$p=-\rho$ without crossing the border between domains.

\section{Conclusions}\label{Conclusions}

We consider the Bianchi type I homogeneous cosmological model with matter with an arbitrary barotropic equation of state and found a system of ordinary differential equations \eqref{eq8}, \eqref{eq9} relating its parameters, such as cosmological time, the rate of expansion along different directions, the density and pressure of the matter filling the universe, including dark matter and dark energy. It give any researcher an opportunity to study the dependence of the parameters of the model without using the mathematical apparatus of the general relativity theory. We provide a solution of this system in the form of integrals \eqref{eq10}, \eqref{eq11} and obtain the generic metric \eqref{eq12}, \eqref{eq13} for this model. In the article it is demonstrated how easy it is to obtain a solution for a known EoS, using the example of the Bianchi type I $\Lambda$CDM homogeneous model.

We analyzed the most general properties of this model, characteristic of any kind of matter that fills the Universe, taking into account two trivial facts: at the some moment, like now, the matter density is positive and the Universe expands. It turned out that it must constantly expand, starting from an anisotropic Kasner-like ``Big Bang'' occurred some finite time ago. There is no ``Big Crunch'' at the end. The expansion either lasts forever or after a finite interval of cosmological time $t$ ends by ``Big Rip''. Other types of future singularities are also possible in the case of special EoS. They are listed in the tables \ref{t1}, \ref{t2} and are analyzed in the section \Sref{future}.

The article introduces a quantitative indicator $W$ of the degree of anisotropy of the considered cosmological model \eqref{an}, \eqref{an1}. The evolution of Bianchi I model is accompanied by isotropization for practically any EoS with a few exceptions, such as the vacuum Kasner solution and some types of future singularities. The Universe filled with matter with reasonable EoS becomes more and more isotropic.

In the inflationary epoch the Universe had experienced very strong isotropization and at the present time its anisotropy is almost imperceptible. This fact allows us to consider homogeneous anisotropic cosmological model as a possible candidate for the description of the observed Universe.

Consideration of other homogeneous cosmological models confirms this conclusion. I plan to write a paper which will present the results of considering isotropization in the case of other types of Bianchi models. 

All this leads us to a conclusion that goes beyond homogeneous models. The observed high degree of isotropy of the Hubble expansion, albeit distorted by cosmic flows, may not be a consequence of the isotropy of its birth. A universe that has gone through an inflationary stage after an anisotropic ``Big Bang'' can provide the same pattern of expansion.

{\bf Data availability statement}

No new data were created or analysed in this study.

{\bf Acknowledgement}

I thank the people and the government of the Swiss  Confederation for supporting Ukrainian scientists in wartime. I thank SwissMAP for funding my visit to the University of  Geneva in Spring 2022, and the Département de Physique  Théorique  and CERN for the opportunity to prolong  this  visit.

{\bf ORCID iD}

Serge Parnovsky https://orcid.org/0000-0002-1855-1404

\end{document}